\documentclass[conference]{IEEEtran}
\usepackage{cite}
\usepackage{graphicx} 
\usepackage{subcaption}
\usepackage{amsmath}
\usepackage{amssymb}
\usepackage{enumitem}

\usepackage[margin=2cm]{geometry}
\usepackage{tabularx}
\usepackage{makecell}
\usepackage{booktabs}
\newcolumntype{Y}{>{\raggedleft\arraybackslash}X}

\title{Generative Decoding of Compressed CSI \\ for MIMO Precoding Design}

\author{
    \IEEEauthorblockN{Hao Luo\IEEEauthorrefmark{1}, Saeed R. Khosravirad\IEEEauthorrefmark{2}, Ahmed Alkhateeb\IEEEauthorrefmark{1}}
    \IEEEauthorblockA{\IEEEauthorrefmark{1}School of Electrical, Computer, and Energy Engineering, Arizona State University}
    \IEEEauthorblockA{\IEEEauthorrefmark{2}Nokia Bell Laboratories}
}

\begin{document}
\bstctlcite{IEEEexample:BSTcontrol}

\maketitle

\begin{abstract}
    Massive MIMO systems can enhance spectral and energy efficiency, but they require accurate channel state information (CSI), which becomes costly as the number of antennas increases. While machine learning (ML) autoencoders show promise for CSI reconstruction and reducing feedback overhead, they introduce new challenges with standardization, interoperability, and backward compatibility. Also, the significant data collection needed for training makes real-world deployment difficult. To overcome these drawbacks, we propose an ML-based, decoder-only solution for compressed CSI. Our approach uses a standardized encoder for CSI compression on the user side and a site-specific generative decoder at the base station to refine the compressed CSI using environmental knowledge. We introduce two training schemes for the generative decoder: An end-to-end method and a two-stage method, both utilizing a goal-oriented loss function. Furthermore, we reduce the data collection overhead by using a site-specific digital twin to generate synthetic CSI data for training. Our simulations highlight the effectiveness of this solution across various feedback overhead regimes.
\end{abstract}

\section{Introduction}
    Multiple-input and multiple-output (MIMO) systems enhance both energy and spectral efficiency by using multiple antennas for beamforming and spatial multiplexing. To achieve this, the base station (BS) requires accurate channel state information (CSI), which is obtained through a downlink CSI acquisition process~\cite{Love2008}. In this process, the BS transmits pilot signals known to the user equipment (UE). The UE then estimates the channel and sends the compressed CSI back to the BS. The BS uses this CSI feedback to design a precoder for data transmission. During the previous decade, massive MIMO~\cite{Larsson2014} has been developed in response to growing demands for higher throughput and more connected users in wireless systems. This involves deploying a larger number of antennas at the BS, but also results in significant feedback overhead. The classical methods like CSI quantization~\cite{Roh2006} may no longer be feasible due to limited uplink resources. This motivates the need for more efficient CSI compression and feedback approaches.

    Recently, machine learning (ML) techniques have been applied to CSI compression and feedback~\cite{Wen2018}, typically using an encoder-decoder architecture. In this setup, the UE uses an ML encoder to compress the estimated channel, and the BS uses an ML decoder to reconstruct it. However, existing encoder-decoder approaches face three main challenges: (i) \textit{Interoperability}: Encoder-decoder designs require joint training and collaboration between the BS and UE vendors, who often consider their datasets and model designs proprietary. (ii) \textit{Backward compatibility}: Integrating these approaches into standards like 3GPP~\cite{3GPP} would require significant changes, making them incompatible with previous versions. (iii) \textit{Data collection overhead}: ML models need large amounts of training data which may not be practical to collect, especially in massive MIMO systems.
    
    To address these challenges, prior work has proposed decoder-only approaches~\cite{Guo2022, Chen2024,Liu2025}. For instance, the authors in~\cite{Guo2022} use a random vector quantization based codebook to compress CSI at the UE, while an ML model at the BS refines the reported beam. In~\cite{Chen2024}, the authors proposed an ML model that can recover CSI, compressed by linear projection at the UE, with an arbitrary compression ratio. A similar approach is also studied in~\cite{Liu2025} but additionally incorporates environmental information into the CSI recovery process via a scene graph. While these approaches show potential to address the interoperability issue, their effectiveness with existing standard methods, such as 3GPP Type-I/II codebooks, remains unexplored. Moreover, their training processes often minimize negative cosine similarity or reconstruction loss, not the end objective of achievable user rate. Finally, the challenge of data collection overhead also remains for these methods.

    In this paper, we propose an ML-based decoder-only solution. This approach is deployed at the BS and uses standardized codebooks as the CSI encoder at the UE, inherently ensuring both interoperability and backward compatibility. Our contributions are summarized as follows:
    \begin{itemize}
        \item  We propose a generative decoder for compressed CSI that produces precoding matrices. This decoder is trained using two proposed approaches, an end-to-end and a two-stage method, with a goal-oriented objective function based on the achievable rate of the UEs.
        \item  To reduce data collection overhead, we propose using a site-specific digital twin to generate synthetic CSI for offline training. The decoder, trained on this data, can then be deployed in a real-world system. We construct a digital twin dataset to study the impact of imperfections.
    \end{itemize}
    Simulation results demonstrate the effectiveness of the proposed generative decoder in CSI refinement. Specifically, the proposed goal-oriented loss outperforms reconstruction loss across different feedback overhead regimes. Furthermore, the decoder trained on digital twin data shows gains compared to the standardized codebook, benefiting from the environmental knowledge provided by the site-specific digital twins.

\section{System and Channel Models}
    We consider a massive MIMO system, where a BS with $N_\mathrm{t}$ antennas communicates with $U$ single-antenna users. The system is assumed to employ an OFDM scheme with $K$ subcarriers. At the $k^{\mathrm{th}}$ subcarrier, the BS precodes the transmitted symbols $\mathbf{s}_k \in \mathbb{C}^{U\times1}$ using linear precoding, and the transmitted signal can be expressed as
    \begin{equation}
        \mathbf{x}_k = \sum_{u=1}^{U} \mathbf{f}_{k,u} s_{k,u} = \mathbf{F}_k \mathbf{s}_{k},
    \end{equation}
    where $s_{k,u} \in \mathbb{C}$ is the data symbol for the $u^{\mathrm{th}}$ user, and the symbols $\mathbf{s}_k \in \mathbb{C}^{U \times 1}$ satisfy $\mathbb{E}[\mathbf{s}_k \mathbf{s}_k^\mathrm{H}] = \mathbf{I}$. $\mathbf{F}_k \in \mathbb{C}^{N_\mathrm{t} \times U}$ is the precoding matrix that contains $U$ precoding vectors, which satisfies the sum-power constraint $\mathrm{Tr}(\mathbf{F}_k \mathbf{F}_k^H) \leq P$. Then, for the $u^{\mathrm{th}}$ user, the received signal can be written as 
    \begin{equation}
        y_{k,u} = \mathbf{h}_{k,u}^\mathrm{H}\mathbf{f}_{k,u} x_{k,u} + \sum_{v \neq u} \mathbf{h}_{k,u}^\mathrm{H}\mathbf{f}_{k,v} x_{k,v} + n_{k,u},
    \end{equation}
    where $\mathbf{h}_{k,u} \in \mathbb{C}^{N_\mathrm{t} \times 1}$ is the frequency-domain channel vector, and $n_{k,u} \sim \mathcal{CN}(0, \sigma^2)$ represents the additive Gaussian white noise. We adopt a wideband geometric channel model, and the delay-d channel $\mathbf{h}_{d,u} \in \mathbb{C}^{N_\mathrm{t} \times 1}$ can be written as
    \begin{equation}
        \mathbf{h}_{d,u} = \sum_{l=1}^{L}\alpha_{l,u} \, p(d \, T_\mathrm{S}-\tau_{l,u})\mathbf{a}(\phi^\mathrm{az}_{l,u}, \phi^\mathrm{el}_{l,u}),
    \end{equation}
    where $p(\tau)$ denotes the pulse shaping function, representing a $T_\mathrm{S}$-spaced signal evaluated at $\tau$ seconds. The channel consists of $L$ paths, and each path contains the complex gain $\alpha_{l,u}$, propagation delay $\tau_{l,u}$, and the azimuth and elevation angles of departure, denoted by $\phi^\mathrm{az}_{l,u}$ and $\phi^\mathrm{el}_{l,u}$. $\mathbf{a}(\phi^\mathrm{az}_{l,u}, \phi^\mathrm{el}_{l,u})$ is the array response vector. Then, the frequency-domain channel vector on the $k^{\mathrm{th}}$ subcarrier can be written as
    \begin{equation}
        \mathbf{h}_{k,u} = \sum_{d=0}^{D-1}\mathbf{h}_{d,u} \exp\left(-j\frac{2\pi k}{K}d\right), 
    \end{equation}
    where $D$ denotes the length of the channel tap. The sum rate of the downlink data transmission can be expressed as 
    \begin{align} \label{eq:sum_rate}
        &R(\{\mathbf{H}_u\}_{u=1}^{U}, \{\mathbf{F}_k\}_{k=1}^{K}) \\ \nonumber
        &= \frac{1}{K} \sum_{k=1}^K \sum_{u=1}^U  \log_2\left(1+\frac{\left|\mathbf{h}_{k,u}^\mathrm{H}\,\mathbf{f}_{k,u}\right|^2}{\sum_{v \neq u} \left|\mathbf{h}_{k,u}^\mathrm{H}\,\mathbf{f}_{k,v}\right|^2 + \sigma^2}  \right),
    \end{align}
    where $\mathbf{H}_u \in \mathbb{C}^{N_\mathrm{t} \times K}$ is the downlink channel matrix of the $u^{\mathrm{th}}$ UE, which is obtained by stacking the channel vectors of the $K$ subcarriers, i.e., $\mathbf{H}_u = [\mathbf{h}_{1,u}, \hdots, \mathbf{h}_{K,u}]$.

\section{Problem Formulation}
    In this paper, our aim is to design a goal-oriented generative decoder for the BS that efficiently decodes and refines the compressed CSI reported by the UE. Specifically, the decoder's output consists of precoding matrices that are optimized to improve the end objective, e.g., the achievable rate of UEs. The overall process of CSI compression and recovery begins with the BS sending pilot signals to the UE. The UE then uses these signals to estimate the channel matrix, denoted as $\widehat{\mathbf{H}}_u$. Next, a standardized encoder $f(\cdot)$, e.g., 3GPP Type I/II codebooks~\cite{3GPP}, is used on the UE to compress the estimated channel. The UE subsequently reports the compressed CSI to the BS. It is worth noting that the CSI feedback process is conducted individually for each UE. Based on the feedback, the generative decoder $g_\theta(\cdot)$, parameterized by $\theta$, outputs the precoding matrices $\{\widehat{\mathbf{F}}_k\}_{k=1}^{K}$. The entire process can be formally expressed as
    \begin{equation}
        \{\widehat{\mathbf{F}}_k\}_{k=1}^{K} = g_\theta(f(\{\widehat{\mathbf{H}}_u\}_{u=1}^{U})).
    \end{equation}
    Ultimately, our objective is to find the optimal decoder, denoted by $\theta^\star$, that maximizes the average sum rate for a given communication environment, which is formulated as
    \begin{align}
        \theta^\star = \underset{\substack{\theta}}{\arg\max} \ \mathbb{E}_{\{\mathbf{H}_u\}_{u=1}^{U} \sim \mathcal{H}} \left[ R(\{\mathbf{H}_u\}_{u=1}^{U}, \{\widehat{\mathbf{F}}_k\}_{k=1}^{K}) \right],
    \end{align}
    where $\mathcal{H}$ denotes the channel distribution.

\section{Proposed Solution}
    \begin{figure*}[t]
        \centering
        \includegraphics[width=0.65\linewidth]{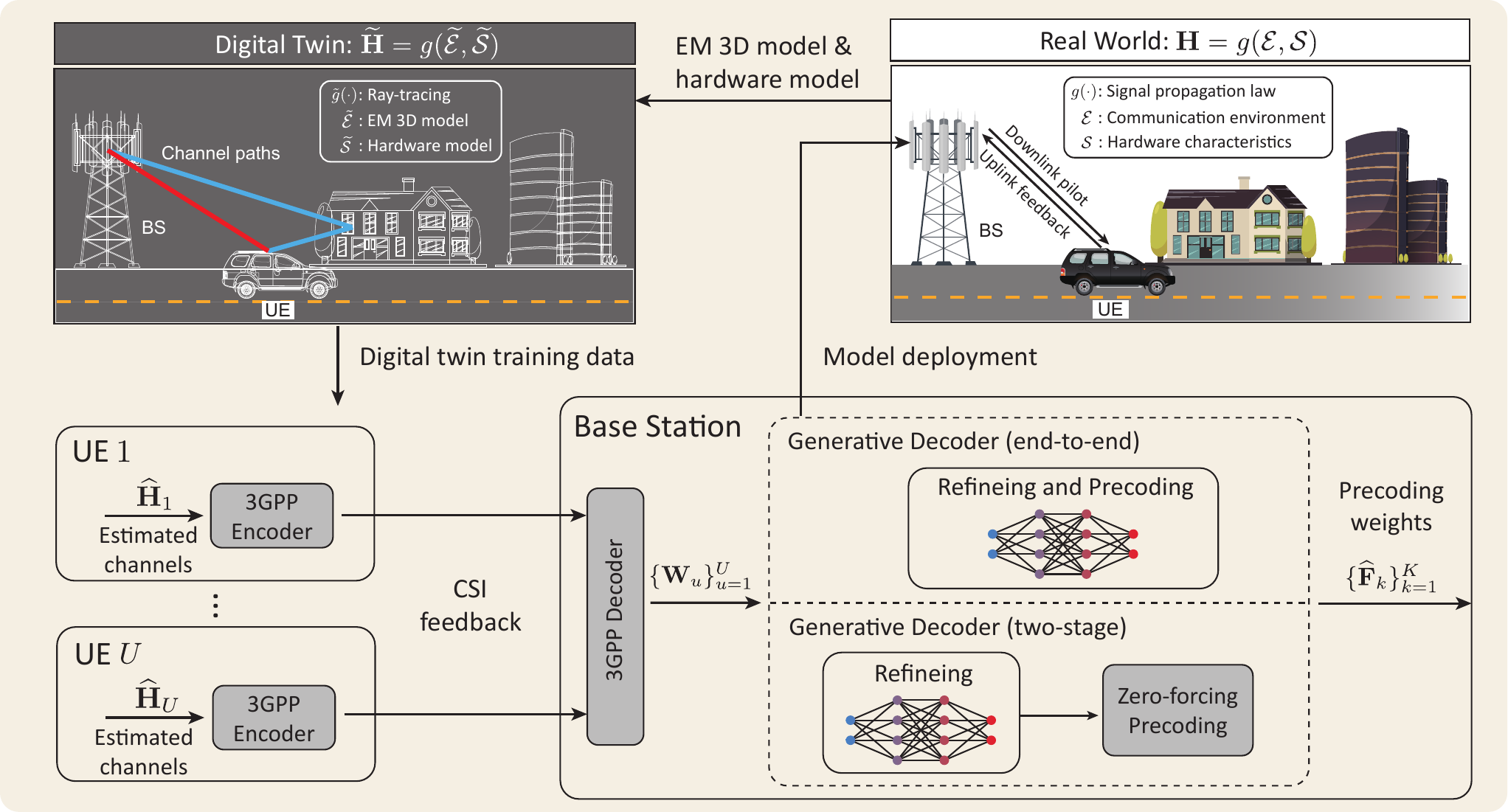}
        \caption{This figure presents the key ideas of the proposed generative decoder solution. The estimated channel is compressed using a standardized encoder on the UE, and a generative decoder refines the reported CSI at the BS. A site-specific digital twin is also leveraged to generate synthetic data for offline training, and the trained generative decoder can be deployed in the real-world system with high performance.}
        \label{fig:key_idea}
    \end{figure*}

    \subsection{Key Idea}
        Existing encoder-decoder models for CSI compression, while efficient, present challenges with interoperability, backward compatibility, and data collection overhead. To overcome these issues, we propose an ML-based, decoder-only CSI feedback solution. Our approach uses a standardized encoder (e.g., 3GPP Type I/II codebooks) on the UE side and a generative decoder at the BS to refine the CSI for precoder design. Our solution offers several key advantages:
        \begin{itemize}
            \item \textit{Interoperability}: As a BS-side solution, it eliminates the need for joint training and enables seamless interoperability between different BS and UE vendors.
            \item \textit{Backward compatibility}: Our method is backward compatible with existing standards since it uses standardized CSI compression, fitting smoothly within current specifications where precoder design is an implementation issue left to the vendor.
            \item \textit{Reduced data collection overhead}: To mitigate the real-world data collection overhead associated with ML-based methods, we propose using site-specific digital twins to generate synthetic CSI data for offline training.
        \end{itemize}
        For training the generative decoder, we propose using a goal-oriented objective function, i.e., the achievable rate of UEs. This approach aims to achieve better training efficiency by directly optimizing for the end objective. The key idea of the propose solutions is illustrated in Fig. \ref{fig:key_idea}. In the following subsections, we will present the proposed solution in detail.

    \subsection{Generative Decoder for Compressed CSI}
        \textbf{Standardized Encoder:} We use 3GPP Type-I/II codebooks as the standardized encoder for UE CSI compression. This process, defined by the 3GPP standard, compresses CSI on a per-subband basis to reduce feedback overhead. A subband consists of multiple physical resource blocks (PRBs), each containing 12 subcarriers. We denote the total number of subbands as $N_\mathrm{SB}$. The core of the Type-I/II codebook is to approximate the estimated channel using a set of discrete Fourier transform (DFT) beams. For a single-polarized antenna array at the BS, the CSI, denoted by $\mathbf{W}_u \in \mathbb{C}^{N_\mathrm{t} \times N_\mathrm{SB}}$, is represented as
        \begin{equation} \label{eq:csi_recon_codebook}
            \mathbf{W}_u = \mathbf{W}_{1, u} \mathbf{W}_{\mathrm{C}1, u} \mathbf{W}_{\mathrm{C}2, u}  \mathbf{\Lambda}_u,
        \end{equation}
        where $\mathbf{W}_{1, u} \in \mathbb{C}^{N_\mathrm{t} \times L}$ contains the $L$ selected wideband DFT beams. $\mathbf{W}_{\mathrm{C}1, u} \in \mathbb{C}^{L \times L}$ is a diagonal matrix with quantized wideband amplitudes. $\mathbf{W}_{\mathrm{C}2, u} \in \mathbb{C}^{L \times N_\mathrm{SB}}$ contains the quantized subband phases and amplitudes for all subbands. $\mathbf{\Lambda}_u \in \mathbb{C}^{N_\mathrm{SB} \times N_\mathrm{SB}}$ is a diagonal matrix with normalization factors for all subbands. The UE feeds back indices of the selected DFT beams along with quantized amplitudes and coefficients.  The specific method for selecting these values is left as an implementation choice in the standard; in this work, we adopt the method described in~\cite{Fu2023}. While higher feedback overhead allows for a more accurate channel approximation, the BS can only reconstruct a lower-resolution version of the channel using the shared codebooks. The main limitation of these standard codebooks is their lack of site-specific environmental knowledge, which constrains the performance of the resulting precoder. This limitation motivates our use of a generative decoder to refine the compressed CSI with site-specific priors.
        
        \textbf{Generative Decoder:} The compressed CSI reported by UEs is in the form of bits. Before refinement with the generative decoder, we first reconstruct this compressed CSI using the shared Type-I/II codebook at the BS, a process we refer to as the \textit{standardized decoder}. This process begins when the BS receives compressed CSI feedback from $U$ users. It then reconstructs the CSI based on \eqref{eq:csi_recon_codebook}, which we denote as $\{\mathbf{W}_u\}_{u=1}^{U}$. Next, we convert the subband-level CSI to subcarrier-level CSI by assigning each subcarrier to its affiliated subband. Our proposed generative decoder refines this subcarrier-level CSI using an ML model. To implement this decoder, we propose schemes: An end-to-end approach and a two-stage approach. These two methods are detailed below.
        \begin{enumerate}[label=(\roman*)]
            \item \textit{End-to-end approach}: First, we propose an end-to-end approach for training the generative decoder. This method enables the decoder to jointly learn how to refine the CSI for each user and design a precoder that effectively mitigates inter-user interference. Based on the subcarrier-level CSI, the generative decoder is trained to produce the precoder matrix for all users and subcarriers, represented by $\{\widehat{\mathbf{F}}_{k}\}_{k=1}^{K}$. The model is trained using a goal-oriented loss function, specifically the negative sum rate. This can be expressed as
            \begin{equation}
                \mathcal{L}_\mathrm{e2e} (\theta) = \frac{1}{N_\mathrm{D}} \sum_{n=1}^{N_\mathrm{D}} - R(\{\mathbf{H}_{n, u}\}_{u=1}^{U}, \{\widehat{\mathbf{F}}_{n, k}\}_{k=1}^{K}),       
            \end{equation}
            where $N_\mathrm{D}$ is the number of data points, and each sample contains the CSI and ground-truth channel for all $U$ users.
            
            \item \textit{Two-stage approach}: We also propose a two-stage approach that offers more flexibility than the end-to-end method, as it is not limited to a fixed number of users. In this approach, each data point contains the CSI and ground-truth channel for a single user. In the first stage, the same decoder is used to individually refine each user's subcarrier-level CSI. We denote this refined CSI as $\widehat{\mathbf{W}}_{u} = [\mathbf{w}_{1,u}, \hdots, \mathbf{w}_{K,u}] \in \mathbf{C}^{N_\mathrm{t} \times K}$. To train the decoder for this stage, we use a goal-oriented loss function based on the negative single-user rate. The single-user rate is expressed as 
            \begin{align}
                R_\mathrm{SU}(\mathbf{H}_{u}, \widehat{\mathbf{W}}_{u}) = \frac{1}{K} \sum_{k=1}^K   \log_2\left(1+\frac{\left|\mathbf{h}_{k,u}^\mathrm{H}\,\mathbf{w}_{k,u}\right|^2}{\sigma^2}  \right).
            \end{align}
            The corresponding loss function can be expressed as
            \begin{equation}
                \mathcal{L}_{\mathrm{ts}, \mathrm{rate}} (\theta) = \frac{1}{N_\mathrm{D}} \sum_{n=1}^{N_\mathrm{D}} -  R_\mathrm{SU}(\mathbf{H}_{n}, \widehat{\mathbf{W}}_{u}).
            \end{equation}
            In the second stage, the precoder matrices, $\{\widehat{\mathbf{F}}_{k}\}_{k=1}^{K}$, are designed using zero-forcing precoding based on the refined CSI. To study the benefits of optimizing directly for a communication objective, we compare it against the reconstruction loss, which is given by
            \begin{equation}
                \mathcal{L}_{\mathrm{ts}, \mathrm{recon}} (\theta) = \frac{1}{N_\mathrm{D}} \sum_{n=1}^{N_\mathrm{D}} \| \mathbf{W}_{n}^\star - \widehat{\mathbf{W}}_{n} \|_\mathrm{F}^2,
            \end{equation}
            where $\| \cdot \|_\mathrm{F}$ denotes the Frobenius norm, and $\mathbf{W}_{n}^\star \in \mathbf{C}^{N_\mathrm{t} \times K}$ is the ground-truth CSI, obtained by normalizing the ground-truth channel.
        \end{enumerate}
    
    \subsection{Digital Twin Aided Generative Decoder}
        The availability of channel data is critical for training a generative decoder, but collecting it can require extensive real-world effort. To overcome this, we propose using site-specific digital twins~\cite{Alkhateeb2023} to generate synthetic channel data that closely resemble the real-world counterpart. This data is then used for offline training, allowing the trained model to be deployed in a real-world system without the need for extensive data collection. A real-world channel is determined by three main factors: The \textit{communication environment} $\mathcal{E}$ which includes the physical properties of the BS, UE, and surrounding objects; the \textit{signal propagation law} $g(\cdot)$ that determines how signals travel through the environment; and the \textit{hardware characteristics} $\mathcal{S}$ of the communication equipment. These factors define the real-world channel as
        \begin{equation}
            \mathbf{H} = g(\mathcal{E}, \mathcal{S}).
        \end{equation}
        Digital twins can approximate each of these factors with specialized modules. For example, a digital twin uses a 3D model with electromagnetic (EM) information $\widetilde{\mathcal{E}}$ to represent the real-world environment. Also, it can employ ray tracing software $\widetilde{g}(\cdot)$ to simulate signal propagation and build a hardware model $\widetilde{\mathcal{S}}$ based on physical measurements. Consequently, the channel modeled by a digital twin can be expressed as
        \begin{equation}
            \widetilde{\mathbf{H}} = \widetilde{g}(\widetilde{\mathcal{E}}, \widetilde{\mathcal{S}}).
        \end{equation}
        This approach allows us to train the generative decoder on a large synthetic dataset and then deploy it in real-world systems, which can significantly reduce the overhead of data collection. A key question is how the imperfections inherent in digital twins affect communication performance. This is a topic we will investigate in our simulations.

\section{Simulation Setup}
    \begin{figure}[t]
        \centering
        \includegraphics[width=0.6\linewidth]{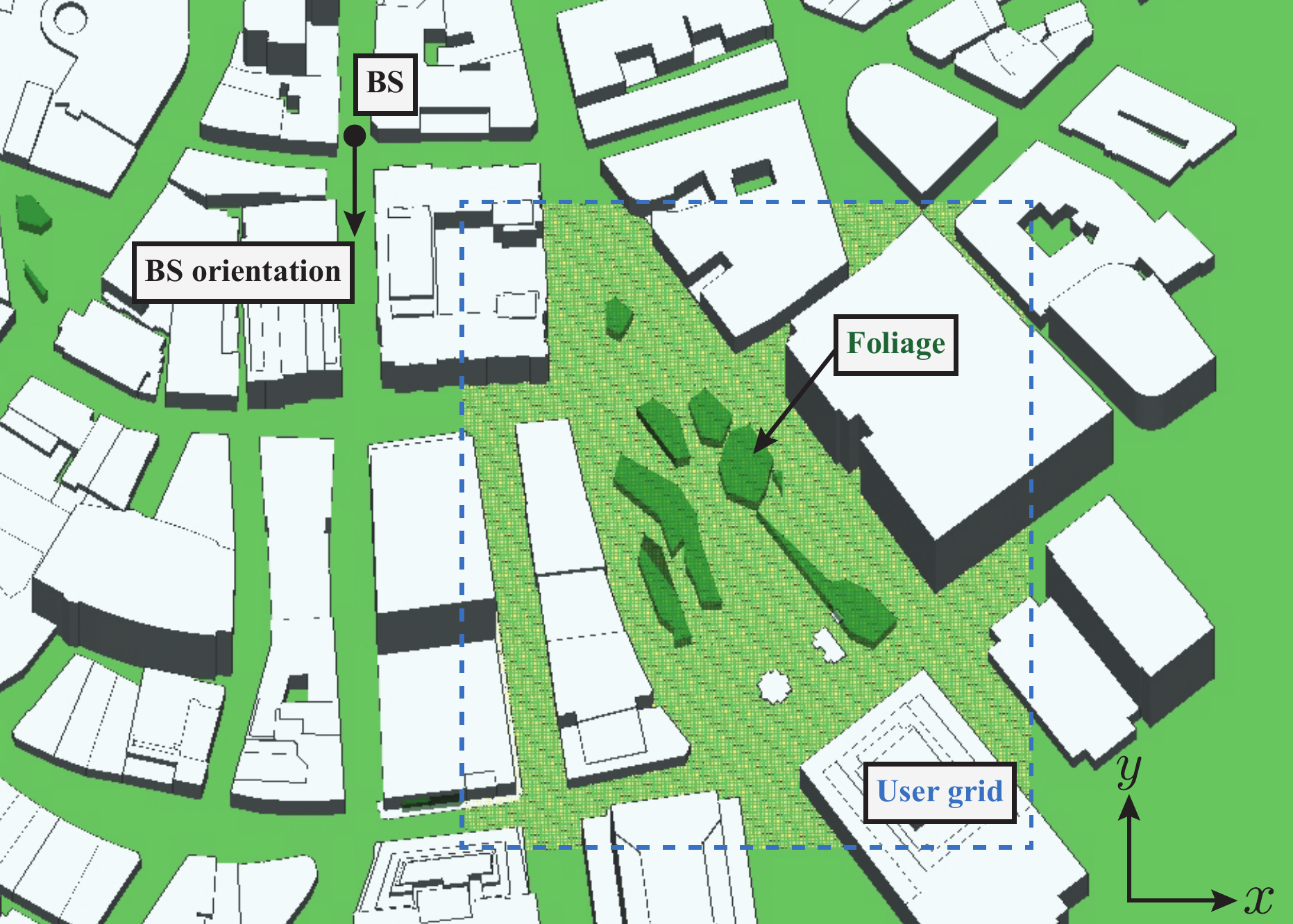}
        \caption{This figure presents a bird's-eye view of the adopted Boston scenario. The BS is oriented toward the negative y-axis. The service area, represented by the user grid, is primarily a non-line-of-sight (NLoS) region} 
        \label{fig:scenario_boston}
    \end{figure}

    \begin{table}[t] 
        \centering
        \caption{System parameters and benchmark method settings}
        \label{tab:parameters}
        \begin{tabularx}{0.95\linewidth}{lcccc}
            \toprule
            \multicolumn{5}{c}{\textbf{System parameters}} \\
            \midrule
            Number of antennas at the BS     &&&& 32  \\
            Number of antennas at the UE     &&&&  1  \\
            Operating frequency              &&&& 3.5~GHz \\
            Number of subcarriers            &&&& 288 \\
            Subcarrier spacing               &&&& 30~KHz  \\
            Equivalent isotropic radiated power at the BS &&&& 30~dBm \\
            Noise figure at the UE           &&&& 7 dB
        \end{tabularx}
        \begin{tabularx}{0.95\linewidth}{lcccc}
            \toprule
            \multicolumn{5}{c}{\textbf{Type-I/II codebook setting}} \\
            \midrule
                    & \makecell{Oversampling\\factor} & \makecell{No. of \\ selected beams} & \makecell{No. of \\ subband} & \makecell{Feedback\\overhead} \\
            \cmidrule(lr){2-5}
            Type-I  &    1 & - & - &    5  \\
                    &    2 & - & - &    6  \\
                    &    4 & - & - &    7  \\
            \cmidrule(lr){2-5}
            Type-II &    4 & 2 &  2 &    23  \\
                    &    4 & 3 &  3 &    47  \\
                    &    4 & 4 &  4 &    77  
        \end{tabularx}
        \begin{tabularx}{0.95\linewidth}{lccccc}
            \toprule
            \multicolumn{6}{c}{\textbf{Non-site-specific datasets}} \\
            \midrule
            DeepMIMO (20 city scenarios) &&&&& $\sim$ 440K samples  \\
            WINNER II                    &&&&& 50K samples \\
            \bottomrule
        \end{tabularx}
    \end{table}

    \begin{figure*}[t]
        \centering
        \begin{subfigure}[b]{0.32\textwidth}
            \centering
            \includegraphics[width=\textwidth]{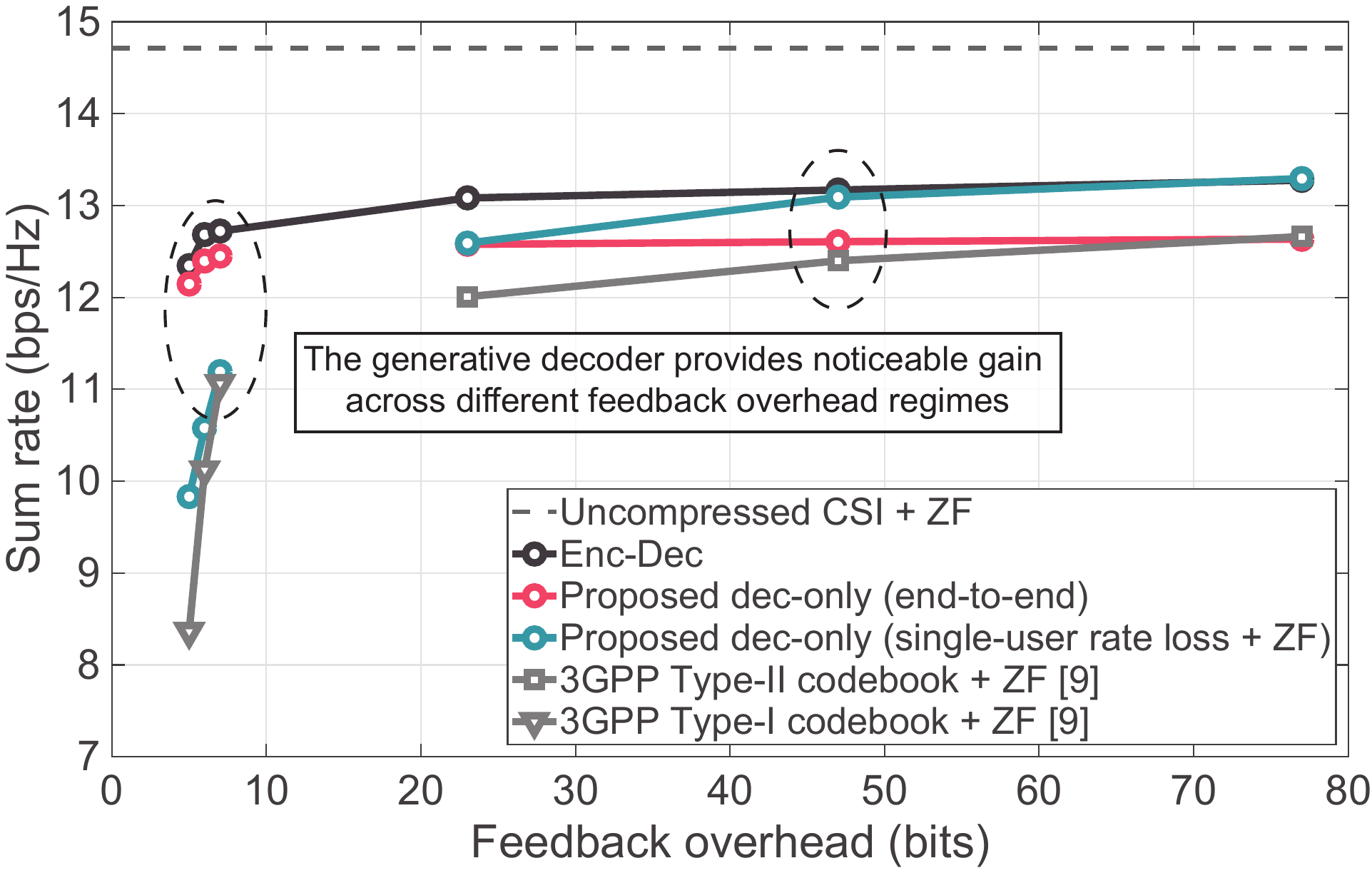}
            \caption{This subfigure presents the comparison between decoder-only and standardized methods.}
            \label{fig:sum_rate_3gpp_codebooks}
        \end{subfigure}
        \hfill
        \begin{subfigure}[b]{0.32\textwidth}
            \centering
            \includegraphics[width=\textwidth]{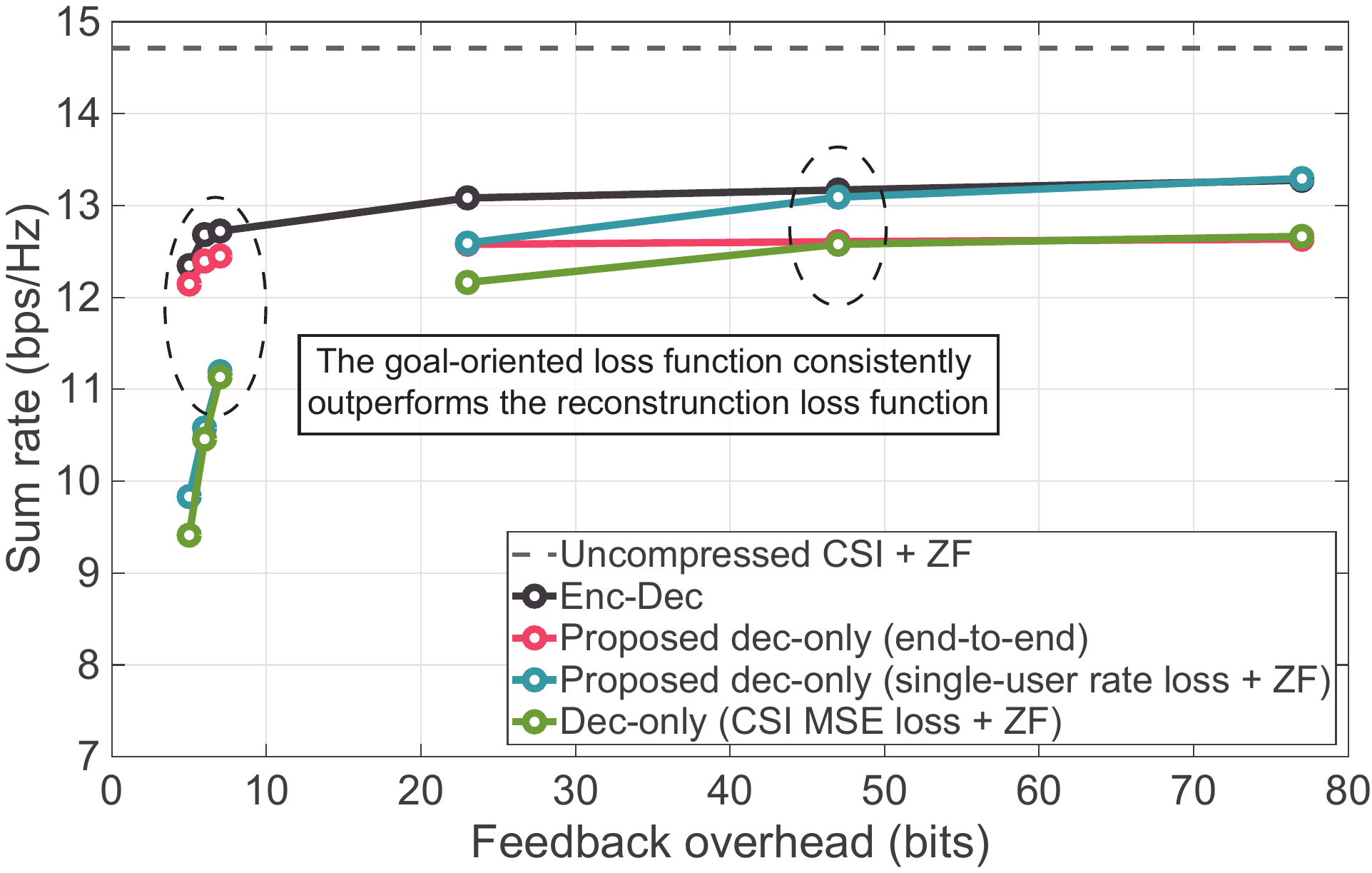}
            \caption{This subfigure presents the comparison between end-to-end and two-stage methods.}
            \label{fig:sum_rate_dec_only}
        \end{subfigure}
        \hfill
        \begin{subfigure}[b]{0.32\textwidth}
            \centering
            \includegraphics[width=\textwidth]{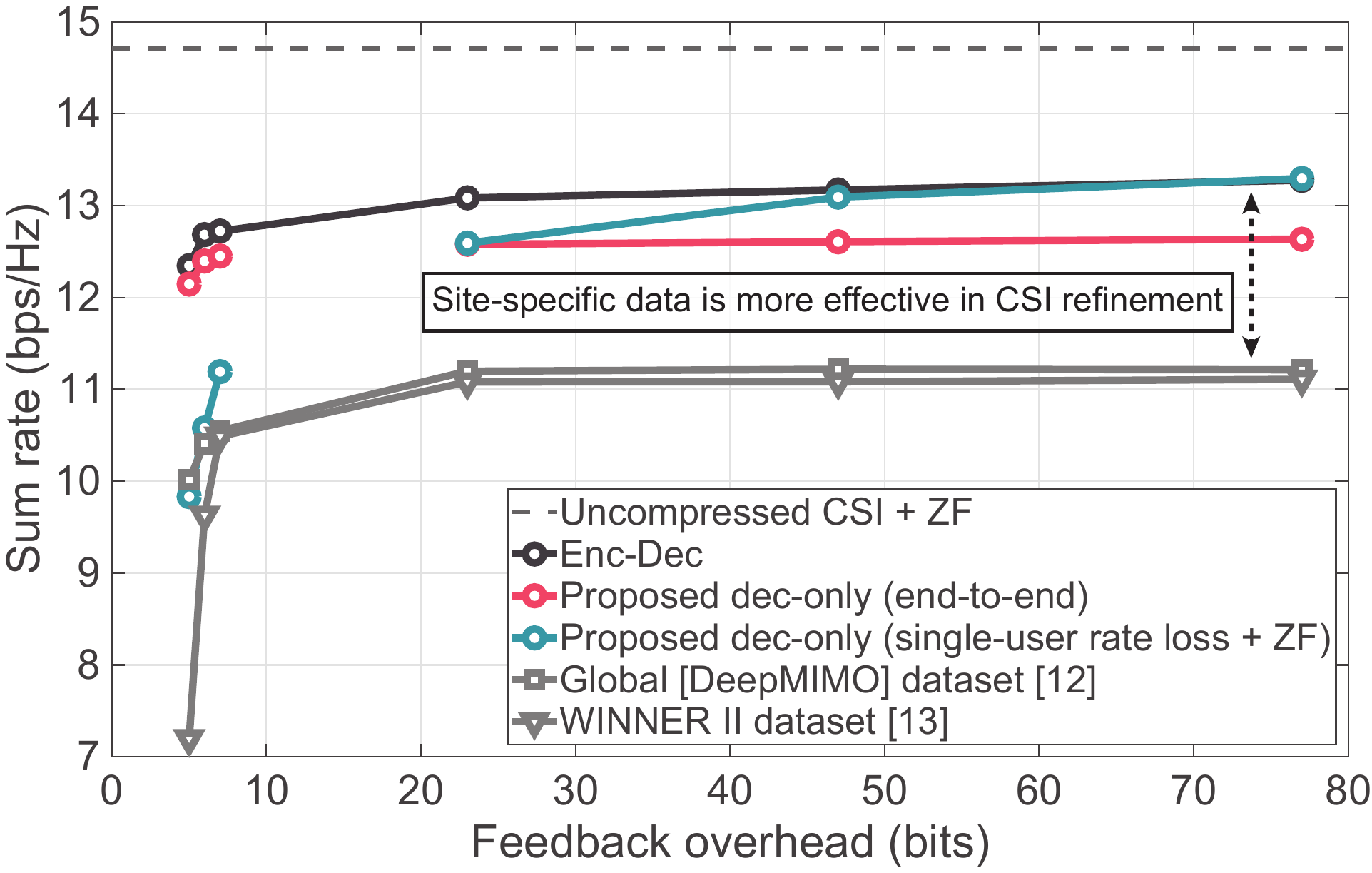}
            \caption{This subfigure presents the comparison between site-specific and non-site-specific data.}
            \label{fig:sum_rate_non_site-specific_dataset}
        \end{subfigure}
        \caption{This figure shows the performance of the generative decoder. The results show its effectiveness by comparing it against codebook-based CSI feedback and a reconstruction loss. The results also emphasize the importance of a site-specific dataset, as shown by the comparison with generic datasets.}
        \label{fig:sum_rate_perfect_dt}
    \end{figure*}
    
    \subsection{Scenario Setup}
        To evaluate the performance of our proposed solution, we use two ray-tracing scenarios in our simulations.
        
        \textbf{Target Scenario}: We first build the target scenario, which serves as our real-world scenario. It is based on downtown Boston and includes detailed models of buildings and foliage, as shown in Fig. \ref{fig:scenario_boston}. In this scenario, the BS is equipped with a 32-antenna uniform linear array and is placed at a height of 15 meters. UEs are placed on a 200m × 230m grid with 0.37m spacing, each at a height of 2m.

        \textbf{Digital Twin Scenario}: The digital twin scenario shares the same basic layout as the target scenario, but it incorporates some practical modeling imperfections. For example, since it is difficult to accurately model the random growth patterns and seasonal changes of trees, we have removed all foliage objects from this scenario. Additionally, to account for potential inaccuracies in real-world data, we introduce building position errors by randomly shifting each building by a certain distance. We will study the impact of these specific imperfections on performance in our simulations 
    
    \subsection{Dataset Generation}
        We adopt the Wireless Insite~\cite{Remcom} ray tracer to simulate signal propagation between the BS and the UEs. This ray tracer identifies the possible propagation paths between each transmitter-receiver pair, providing complex gain, propagation delay, and angles of departure for each path. This information is then used to construct the channel matrix $\mathbf{H}$, with the help of the DeepMIMO channel generator~\cite{Alkhateeb2019}. In our ray-tracing simulations, we set the maximum number of reflections for any given path to four. The operating frequency is 3.5 GHz, with 288 subcarriers and a 30 kHz spacing. For EM materials, the buildings are modeled as concrete and the terrain as wet earth, using the standard ITU parameters for 3.5 GHz. The foliage is treated as an attenuating material with a coefficient of 1 dB per meter. In total, this scenario results in 41,120 data points. These data points are then partitioned into training and testing sets with an 80\% to 20\% ratio, respectively.
    
    \subsection{Deep Learning Model Architecture}
        The generative decoder uses five sequential refining networks to process and refine the input CSI. The decoder's input is a concatenated tensor of CSI from users. Specifically, the real and imaginary parts of the CSI for each of the $U$ users are treated as separate channels, resulting in a total of 2$U$ input channels. This input is processed by the first refining network. Each refining network is a convolutional neural network (CNN) with three layers. The first two layers increase the channel count from 2$U$ to 16, then to 32, using a Leaky ReLU activation function. The final layer reduces the channels back to 2$U$ and uses a Tanh activation function. A skip connection is a key feature, adding the original input to the output of the convolutional layers to preserve information and enhance training stability. In the end-to-end approach, the decoder processes CSI from all users simultaneously, enabling the network to jointly learn CSI refinement and inter-user interference mitigation. The final output is then restructured to form the precoding matrices for all users. In the two-stage approach, the input and output are reduced to process data for a single user at a time.

    \subsection{Benchmarks}
        \textbf{Type-I/II Codebooks}: We use the performance of Type-I/II codebooks as a lower bound. Since our method refines the CSI generated by these codebooks, their performance represents the minimum level we expect to achieve.

        \textbf{Non-Site-Specific Dataset}: To demonstrate the benefits of our site-specific digital twin data, we compare our results with two benchmark datasets. The first is a global dataset containing 20 different city ray-tracing scenarios from the DeepMIMO dataset~\cite{Alkhateeb2019}. The second is a generic statistical dataset generated by the WINNER II channel model~\cite{WinnerII}, for which we use the "Urban Macro Cell" scenario. This comparison highlights the performance gains of training on data that accurately reflects the deployment environment.

        \textbf{Encoder-Decoder Approach}: We also compare our decoder-only method to an encoder-decoder approach, which serves as our empirical upper bound. In this benchmark, the UE uses an ML-based encoder to compress the channel. This encoder is a CNN with six sequential convolutional blocks, each followed by batch normalization and a Leaky ReLU activation. After convolutional layers, the data is flattened and compressed by a fully connected network into an encoded vector. A straight-through estimator approximates the non-differentiable quantization~\cite{Sohrabi2021}, enabling end-to-end training and a binary-like output. We can adjust the feedback overhead by changing the dimension of this encoded vector.

\section{Evaluation Results}    
    We evaluate our proposed generative decoder for compressed CSI using the parameters listed in Table \ref{tab:parameters}. Our evaluation is carried our in two stages using estimated CSI from the UE as data. First, we establish a reference point for effectiveness by training and testing the decoder on data from the target scenario. Next, we evaluate a more practical case where the decoder is trained on synthetic data from the digital twin but tested on data from the target scenario. In both cases, the UE's estimated channel is first compressed using Type-I/II codebooks and then passed to the generative decoder, which produces precoding matrices for two users. The performance is measured by the sum rate calculated using \eqref{eq:sum_rate}. For comparison, we include a genie-aided upper bound, which represents the sum rate achieved with estimated channel and zero-forcing precoding, and an encoder-decoder approach that serves as an empirical upper bound. In the figures, some curves are segmented to differentiate between Type-I and Type-II codebooks. The lower feedback overhead corresponds to Type-I, while the higher feedback overhead corresponds to Type-II. A higher feedback overhead generally leads to better performance.

    \textbf{Can the generative decoder effectively refine codebook-based CSI?}
    Fig. \ref{fig:sum_rate_perfect_dt}(\subref{fig:sum_rate_3gpp_codebooks}) compares our decoder-only solution with the performance of Type-I/II codebooks. The results show that the generative decoder can effectively refine the codebook-based CSI, providing a noticeable gain across different feedback overhead regimes. Specifically, in the low feedback regime (using a Type-I codebook), the end-to-end approach performs better, as the low-resolution channel information from the Type-I codebook makes it difficult to individually refine the CSI for each user. In contrast, the end-to-end approach directly optimizes the precoder and sum rate, ensuring inter-user interference is effectively mitigated. In the high feedback regime (using a Type-II codebook), the two-stage method with a goal-oriented loss function shows superior performance. It can even approach the performance of the encoder-decoder solution. This is because the Type-II codebook provides higher-resolution CSI, which allows the two-stage method to perform individual CSI refinement more accurately.
    
    \textbf{What are the advantages of goal-oriented loss function and site-specific training dataset?} 
    In Fig. \ref{fig:sum_rate_perfect_dt}(\subref{fig:sum_rate_dec_only}), we present the results for  the end-to-end method and two-stage method with two different loss functions. Compared to the reconstruction loss, the approaches using a goal-oriented loss function consistently achieve better performance, which demonstrates the effectiveness of optimizing directly for the end objective. To highlight the advantage of using site-specific data, we compare our solution with an encoder-decoder method trained on non-site-specific datasets in Fig. \ref{fig:sum_rate_perfect_dt}(\subref{fig:sum_rate_non_site-specific_dataset}). We observe that models trained on either a global dataset or a generic statistical dataset do not generalize well to an unseen scenario in the context of CSI refinement. This underscores the critical role of environmental knowledge provided by a site-specific dataset. 

    \begin{figure}[t]
        \centering
        \includegraphics[width=0.7\linewidth]{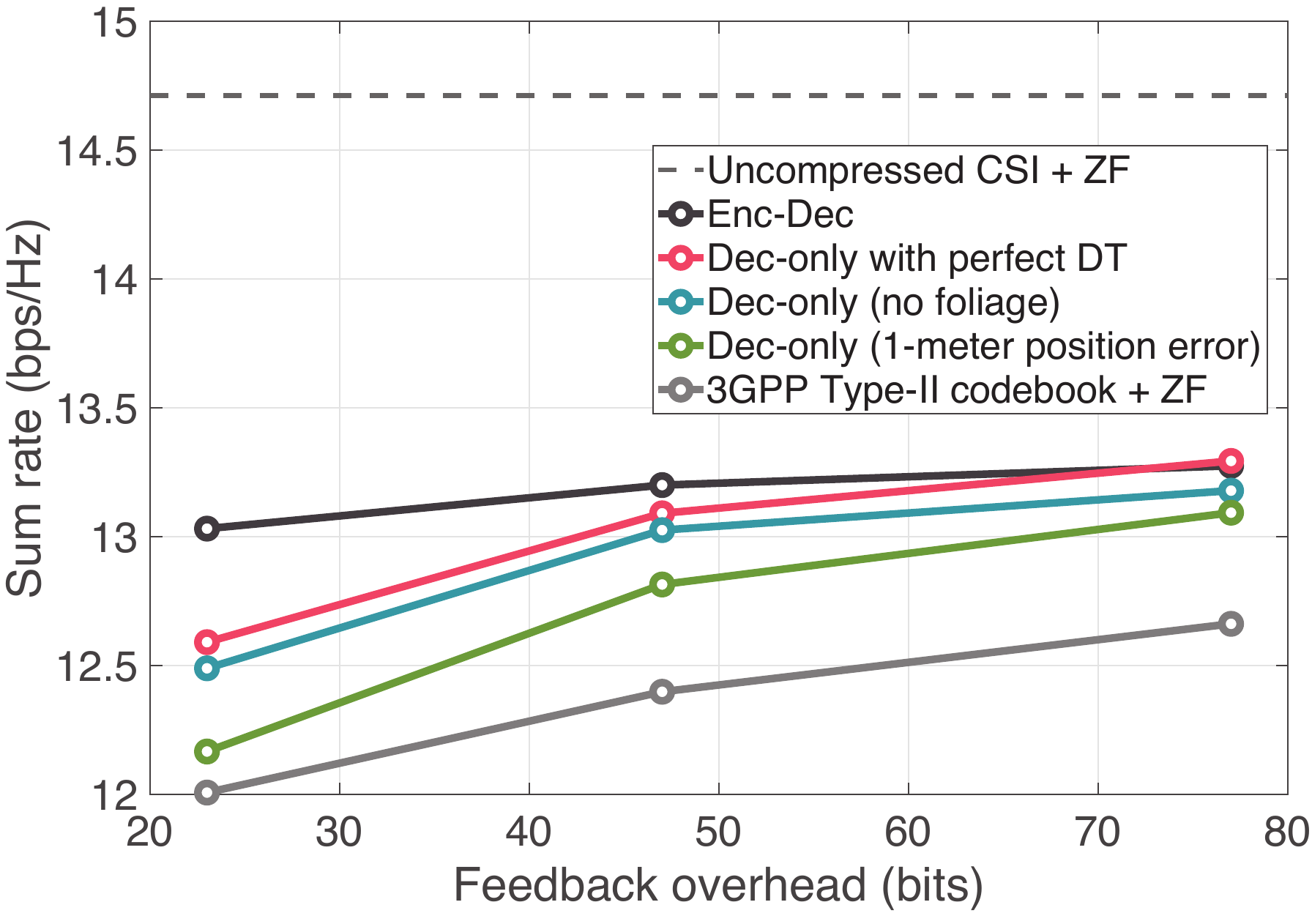}
        \caption{The figure presents results for the two-stage generative decoder, which was trained with a goal-oriented loss function on the digital twin dataset and tested in the target scenario. Despite performance degradation from the digital twin's imperfections, our proposed approach still shows a significant gain over standard Type-II codebooks.}
        \label{fig:sum_rate_imperfect_dt}
    \end{figure}
    
    \textbf{How does digital twin imperfection affect the performance of the generative decoder?} Fig. \ref{fig:sum_rate_imperfect_dt} presents the results for a two-stage generative decoder trained on the digital twin dataset with a goal-oriented loss function, and then tested on the target scenario. We observe that imperfections in the digital twin can cause performance degradation. For instance, building position errors can cause more severe degradation than removing foliage. Despite this, the proposed approach still demonstrates a noticeable gain compared to the standard Type-II codebook. This highlights the feasibility of using synthetic data for training and then deploying the trained model in real-world conditions. Furthermore, an interesting future direction would be to collect a small amount of real-world data to further refine the generative decoder and reduce the performance gap.

\section{Conclusion}
    To overcome the interoperability, backward compatibility, and data overhead challenges in ML-based CSI feedback, we propose a decoder-only solution. In this approach, the UE uses a standardized codebook for CSI compression, while the BS employs a generative decoder to refine the reported CSI for precoder design. To train the decoder, we use a goal-oriented loss function that directly optimizes the achievable rate of the UEs for the end objective. Our simulation results show that the model trained with this goal-oriented loss outperforms benchmark methods across various overhead regimes. Furthermore, the effectiveness in CSI refinement demonstrates the value of incorporating environmental knowledge from the digital twin.

\section*{Acknowledgement}
	This work was supported in part by the National Science Foundation under Grant No. 2426906.


\end{document}